\documentclass[prl,twocolumn,showpacs,superscriptaddress]{revtex4-1}

\usepackage{graphicx}
\usepackage{color}
\usepackage{amsmath}
\usepackage{amssymb}

\usepackage{mathptmx}
\usepackage{times}

\newcommand{\com}[1]{} 

\begin{document}


\title{Rigid cluster decomposition reveals criticality in frictional jamming}
\author{Silke Henkes}
\affiliation{Institute of Complex Systems and Mathematical Biology, Department of Physics, University of Aberdeen, Aberdeen, Scotland, United Kingdom}
\email{shenkes@abdn.ac.uk}
\author{David A. Quint}
\affiliation{Department of Bioengineering, Stanford University and Department of Plant Biology, Carnegie Institute of Washington, Stanford CA, United States}
\author{Yaouen Fily}
\affiliation{Martin Fisher School of Physics, Brandeis University, Waltham MA, United States}
\author{J. M. Schwarz}
\affiliation{Department of Physics, Syracuse Biomaterials Institute, Syracuse University, Syracuse NY, United States}
\email{jschwarz@physics.syr.edu}

\begin{abstract}
We study the nature of the frictional jamming transition within the framework of rigidity percolation theory.  Slowly sheared frictional packings are decomposed into rigid clusters and floppy regions with a generalization of the pebble game including frictional contacts. We discover a second-order transition controlled by the emergence of a system-spanning rigid cluster accompanied by a critical cluster size distribution. Rigid clusters also correlate with common measures of rigidity. We contrast this result with frictionless jamming, where the rigid cluster size distribution is noncritical.  
\end{abstract}


\maketitle


The interplay of constraints, forces, and driving gives rise to the jamming transition in granular media. It is now well-established that the frictionless jamming transition has characteristics of both second- and first-order transitions. Both the average coordination number and the largest rigid cluster size jump at the transition, yet there exists a diverging lengthscale~\cite{OHern03,Ellenbroek15,Wyart05,Goodrich13}. Frictional jamming is more puzzling:
The hysteresis observed in the stress-strain rate curves of stress-controlled flow simulations~\cite{Otsuki10,Heussinger13,ciamarra_jamming_2009,Grob14} and experiments~\cite{Bi11} has lead to an interpretation as a first-order transition.
Yet, signs of second-order criticality appear when treating the fraction of contacts at the Coulomb threshold as an additional parameter~\cite{Shundyak07, Henkes10, avalanche10}.

\begin{figure}[ht]
 \centering
 \includegraphics[width=0.85\columnwidth,trim=0mm 0mm 0mm 0mm,clip]{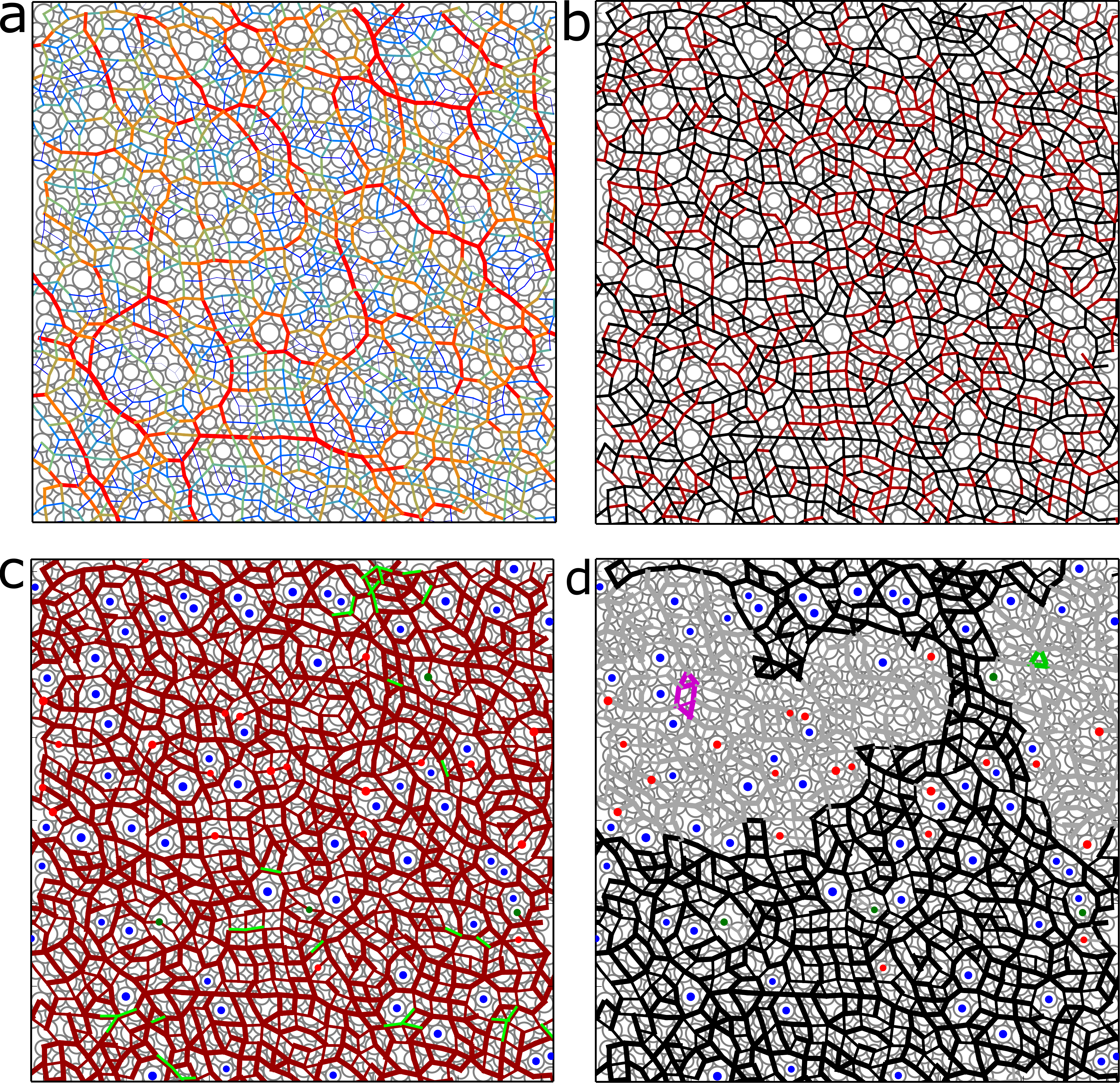}
 \caption{The pebble game and rigidity percolation for a sample $N=1024$ frictional packing. \emph{a} - Force chains, with anisotropic stresses due to Lees-Edwards boundary conditions. \emph{b} - Contact network, with frictional (black) and sliding contacts (red). \emph{c} - Result of the $(3,3)$ pebble game where frictional and sliding contacts have been mapped to double (thick)  and single (thin) bonds, respectively. Red are pebble-covered bonds, green are overconstrained bonds, and colored disks are leftover pebbles. \emph{d} - Rigid cluster decomposition: This packing is partially rigid and consists of three rigid clusters (black, purple and green) and regions of floppy bonds (gray). }
 \label{fig:simulated_game}
\end{figure}

To elucidate the frictional jamming transition from a microscopic viewpoint, 
we extend concepts and tools from rigidity percolation, i.e., the onset of mechanical rigidity in disordered spring networks~\cite{DeGennes76,Feng84,Jacobs95,Moukarzel95}, to frictional packings.
The former is driven by the emergence of a system-spanning rigid cluster that can be mapped out (in 2d) using the pebble game~\cite{Jacobs97}, an improved constraint counting method that goes beyond mean-field by identifying redundant constraints.
We, for the first time, implement a generalized pebble game for 2d frictional systems and use it to identify rigid clusters in very slowly sheared packings. 
As we show below, this allows us to identify a second-order rigidity transition and to link stresses and nonaffine motions to the microscopic structure of frictionally jammed packings.

\emph{Generalized isostaticity:} To establish context, we first review the application of Maxwell constraint counting to jamming~\cite{Maxwell1864}. For $N$ particles in $d$ dimensions and a mean number of contacts per particle $z$, interparticle forces yield $dzN/2$ constraints. Since each particle has $\frac{1}{2}d(d+1)$ translational and rotational degrees of freedom, there are $\frac{1}{2}(N-1)d(d+1)$ total degrees of freedom (subtracting out global degrees of freedom). When these match the force constraints, we arrive at the isostatic criterion, or  $d z N/2 = \frac{1}{2}(N-1)d(d+1)$. In the limit $N\rightarrow \infty$, $z_{\text{iso}}= d+1$ for frictional granular materials. For frictionless packings, we ignore rotations and obtain the familiar $z_{iso}=2d$.

Despite being mean field, i.e. neglecting spatial correlations, isostaticity works seemingly well to locate the jamming transition in static frictionless systems~\cite{OHern03}. For frictional systems, however, numerical and experimental evidence point to a range $d+1<z<2d$ at the transition, with a matching density range from random loose packing~\cite{onoda_liniger} to random close packing. To resolve this conundrum, a \emph{generalized isostaticity} criterion was introduced \cite{Shundyak07, Henkes10}, that accounts for contacts at the Coulomb friction threshold providing one less constraint:
\begin{equation} z_{\text{iso}}^{m}= (d+1) + 2 n_m /d , \label{eq:gen_ziso}\end{equation}
where $n_m$ is the mean number of such fully mobilized contacts per particle. Equation \eqref{eq:gen_ziso} describes a \emph{line} of transition points interpolating from $z=d+1$ at $n_m=0$, corresponding to the friction coefficient $\mu = \infty$ limit, to $n_m=1$ at $z_{\text{iso}}=4$, corresponding to the $\mu = 0$ limit~\cite{Shundyak07}.
\begin{figure*}[t]
\includegraphics[width=0.75\textwidth]{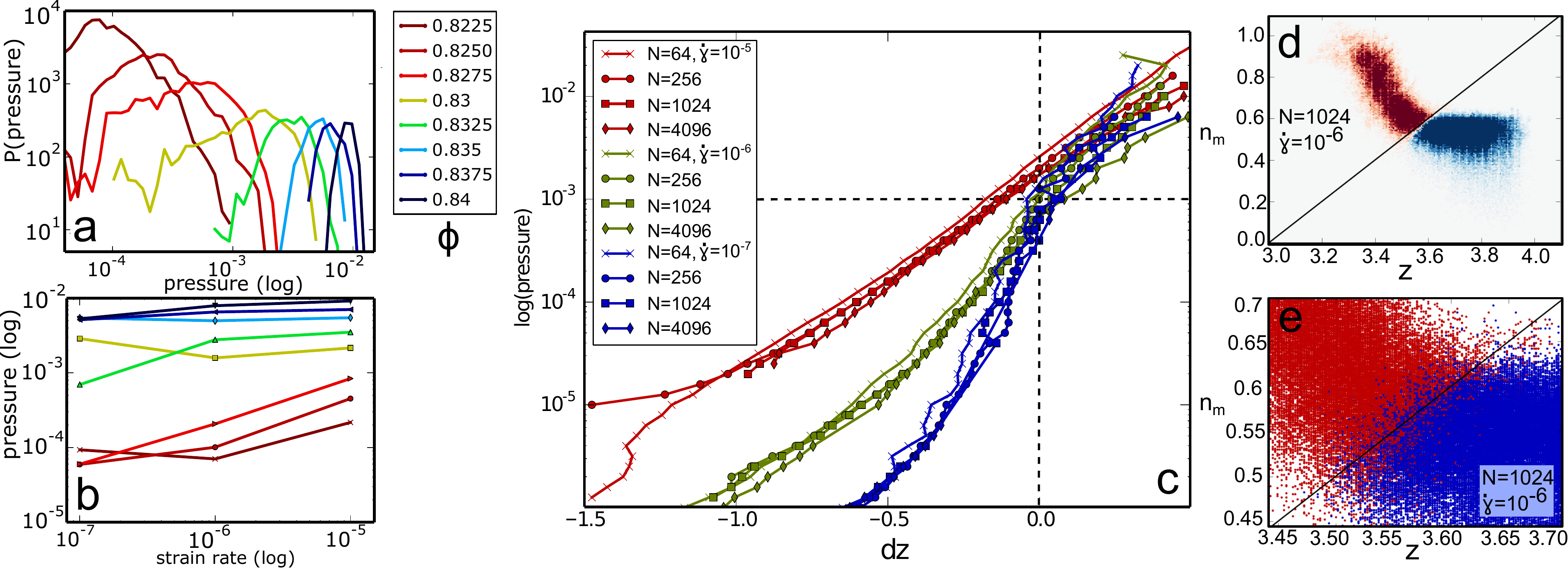}
\caption{ \emph{a} - Pressure distributions for 8 densities $\phi = 0.825-0.84$ accross the frictional jamming transition, for $N=1024$ and $\dot{\gamma}=10^{-5}$. \emph{b} - Stress-strain relations for the same densities and $N$, and strain rates $\dot{\gamma}=10^{-5}-10^{-7}$. \emph{c} - Correlation between distance from generalized isostaticity $dz=z-(3+n_m)$ and pressure (log-scale). \emph{d} - Generalized isostaticity: Probability density of the system in $(n_m,z_{iso})$ phase space, combined from densities $\phi= 0.825-0.845$; $N=1024$ and $\dot{\gamma}=10^{-6}$. The red area has $p<10^{-3}$ and is unjammed, while the blue region has $p>10^{-3}$ and is jammed. \emph{e} - Fluctuations for the same data set, each dot is a jammed (blue) or an unjammed (red) packing. }
 \vspace{-3mm}
\label{fig:Genziso_panel}
\end{figure*}

\emph{Simulation.} 
To obtain frictional packings near jamming, energy-minimization is not an option for the non-conservative frictional interaction. Instead, we implement a common experimental and simulation protocol, simple shear at strain rate $\dot\gamma$ in Lees-Edwards boundary conditions. As we decrease the strain rate, we expect to reach a ``quasistatic'' limit, where the system transitions through a series of (nearly) force-balanced states so that the mechanical properties of these states are relevant, and we  can use the tools of rigidity percolation. 

We simulate systems of $N=64-4096$ polydisperse particles in two dimensions interacting according to the Cundall-Strack law \cite{Cundall-Strack}. In addition to harmonic purely repulsive normal forces $f_n = k_n \delta$, ($\delta$ is the particle overlap), it comprises an incremental tangential force $df_{\text{tan}}= k_t d t$, where $dt$ is the amount of tangential sliding since the establishment of the contact. The friction force is constrained by the Coulomb criterion $|f_t |\leq \mu f_n$; once the threshold is reached, it continues to slide at $|f_t|=\mu f_n$ until the direction reverses. The energy dumped into the system by shearing is dissipated through linear viscous damping forces, $\mathbf{f}^{\zeta}_{ij}=-\zeta(\mathbf{v}_i - \mathbf{v}_j)$, and a small amount of rotational individual damping. 
We work in scaled units with mean particle radius $\langle r \rangle =1$ and unit stiffness $k_n = k_t =1$. Most of the results below are in the low-friction $\mu=0.1$, low-damping $\zeta=0.1$ limit, for $N=1024$ particles except where specified otherwise. We report additional results for high $\mu=1$ and high $\zeta=1$ in Figures S4 and S5, respectively. Systems are strained for $T=10^6$ time units in all cases, corresponding to a strain of $10$ system lengths for $\dot{\gamma}=10^{-5}$ strain rate, and $1$ system length for $\dot{\gamma}=10^{-6}$. 

\emph{Rigidity percolation.}  We decompose packings into rigid clusters using a $(k=3,l=3)$ pebble game~\cite{Jacobs97}: First, we associate a pebble with each of the $k=3$ degrees of freedom of a particle. 
We then build a constraint network based on the contact network where a contact imposing $n$ constraints translates to $n$ bonds, i.e., fully mobilized contacts with $|f_t|= \mu f_n$ yield one bond while contacts with $|f_t|< \mu f_n$ yield two (Fig.~S1).
 We then recursively move these pebbles along bonds in the constraint network to assign pebbles to bonds, where each pebble-covered bond is equivalent to a degree of freedom being independently constrained. We do so until $l=3$ or more pebbles cannot be assigned. Finally, we map out the rigid clusters for each contact network. This algorithm is an extension of the $(2,3)$ pebble game used in the frictionless case. Please see the SI and Fig.~S2 for more details.

In Fig. \ref{fig:simulated_game}, we show the four stages of moving from the simulated packing to the rigid cluster decomposition. The force chains in Fig. 1a correspond to the constraint network in Fig. 1b and 1c, with frictional contacts (double bonds) in black, and sliding contacts (single bonds) in red. This network forms the basis for the $(3,3)$ pebble game in Fig.1c. In our example, it is possible to assign pebbles to most of the bonds in the network without overconstraining it. In agreement with the beyond-mean field nature of Laman's theorem (see SI), a number of leftover pebbles also remain, especially at rattlers buth also at connected particles (colored circles). Finally, in Fig. 1d, we decompose the system into rigid clusters, that is connected regions where no more than the 3 pebbles linked to the global translations and rotations can be found. We find three rigid clusters including a large, system-spanning one. The remaining bonds are floppy, i.e. not rigid with respect to any of their neighbouring bonds. Our example system is globally rigid, in spite of an average $z$ below the generalized isostaticity criterion. We show that such configurations are generic below. 

\begin{figure}
 \centering
  \includegraphics[width=0.99\columnwidth,trim=0mm 0mm 0mm 0mm,clip]{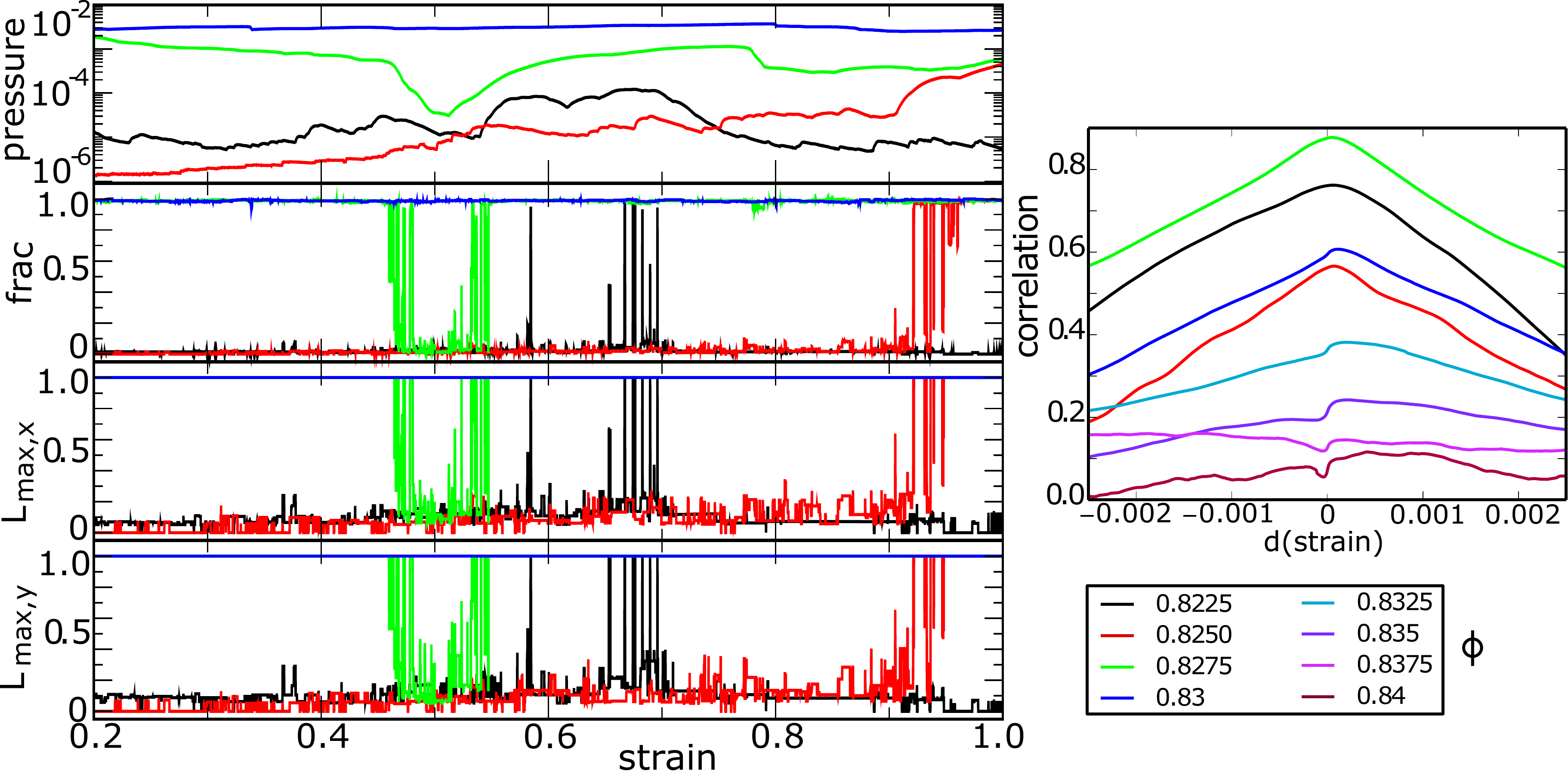}
 \caption{Temporal correlation between rigidity and stresses. Left: Pressure trajectories for four densities close to jamming (log-scale), followed by the fraction of the system that is rigid, (frac), the $x$-length of the largest cluster and the $y$-length of the largest cluster. Right: Time correlation between pressure and rigidity $\langle \text{frac}(t) \log p (t +dt)\rangle$.}
 \vspace{-5mm}
 \label{fig:time_dependence}
\end{figure}

{\it Results:} We first address how the global properties of the system, including mean stresses and distance from isostaticity, depend on density, strain rate and system size (Fig. \ref{fig:Genziso_panel}). Experimental~\cite{Bi11} and simulated frictional systems ~\cite{ciamarra_jamming_2009,Otsuki10,Heussinger13,Grob14} report a hysteresis loop in the stress-strain relations, through a protocol that includes either a strain rate ramp or constant stress driving. As a function of density, $\phi=0.8225-0.84$, we see a transition through jamming, as evidenced by the pressure distribution shifting from a peak at $10^{-4}$ (in units of overlap) to a peak at $10^{-2}$ (Fig. 2a; in Fig. S3 we show that $p$ and $\sigma_{xy}$ are equivalent). Since we perform a constant strain rate simulation, we do not observe hysteresis. We instead find intermittent flips between jammed and unjammed states for the intermediate density samples, in particular $\phi=0.8275$ and $\phi=0.83$ (see Fig.~\ref{fig:time_dependence}). 
When lowering strain rates, a gap in pressure opens between low and high densities, consistent with an approximately Bagnold scaling $p\sim \dot{\gamma}^{1/2}$ dominated by viscous damping forces below jamming, and the appearance of a yield stress above jamming (Fig. \ref{fig:Genziso_panel}b).

We focus our attention on the tuning parameter $dz = z  - z_{iso}^m$, i.e. the distance from  generalized isostaticity. For each strain rate, all of the data for different packing fractions and system sizes collapses onto a unique curve $p(dz)$ (Fig. 2c). For the two lower strain rates, we see two regimes -- a rapid drop below $dz \approx 0$ which depends on $\dot{\gamma}$, separated from a more gradual, universal increase at $dz>0$. The small, positive values of the pressure for $d z<0$ strongly depend on $\zeta$ (see Fig. S5), indicating again that viscous damping forces dominate this regime, consistent with recent results for shear thickening in suspensions \cite{Wyart14}. When we visualize our system in a two dimensional $n_m - z$ phase diagram (Fig.~\ref{fig:Genziso_panel}d), jammed (blue) states defined here by $p\geq 10^{-3}$ exist predominantly in the stable region of the phase diagram below the stability line, while unjammed (red) states mostly exist in the unstable region. There are however significant fluctuations in the transition region with some jammed states below the stability line, suggesting that this mean-field criterion is insufficient (Fig.~\ref{fig:Genziso_panel}e). System trajectories have roughly equal fluctuations in $z$ and $n_m$, unlike the avalanching system of \cite{avalanche10}, where fluctuations along $n_m$ were more prominent, showing that the nature of the trajectories depends on the type of driving.

We now present the results of the rigid cluster decomposition using the $(3,3)$ pebble game. To demonstrate the structural importance of the rigid clusters, we first correlate the time series of rigidity and pressure in the region where we observe intermittent behaviour (see Fig. \ref{fig:time_dependence}). In the second, third and fourth panel, respectively, we show the fraction frac of the bonds belonging to a rigid cluster, the $x$-extent $L_x$ and the $y$-extent $L_y$ of the largest cluster (normalized by system size $L$). Rigid systems are characterized by $\text{frac} \approx 1$, and system spanning clusters in the $x$ and $y$ directions. All three measures correlate with pressure and with each other, though switches between globally rigid and floppy states are significantly faster than pressure changes.  On the right (Fig. 3b), the correlation function $\langle \text{frac}(t) \log p (t +dt)\rangle$ shows strong, symmetric correlations for the mostly unjammed runs $\phi=0.8225-0.83$, and a slight asymmetry indicating that pressure follows rigidity, for the jammed runs $\phi=0.8325-0.84$.

\begin{figure}
 \centering
  \includegraphics[width=0.99\columnwidth,trim=0mm 0mm 0mm 0mm,clip]{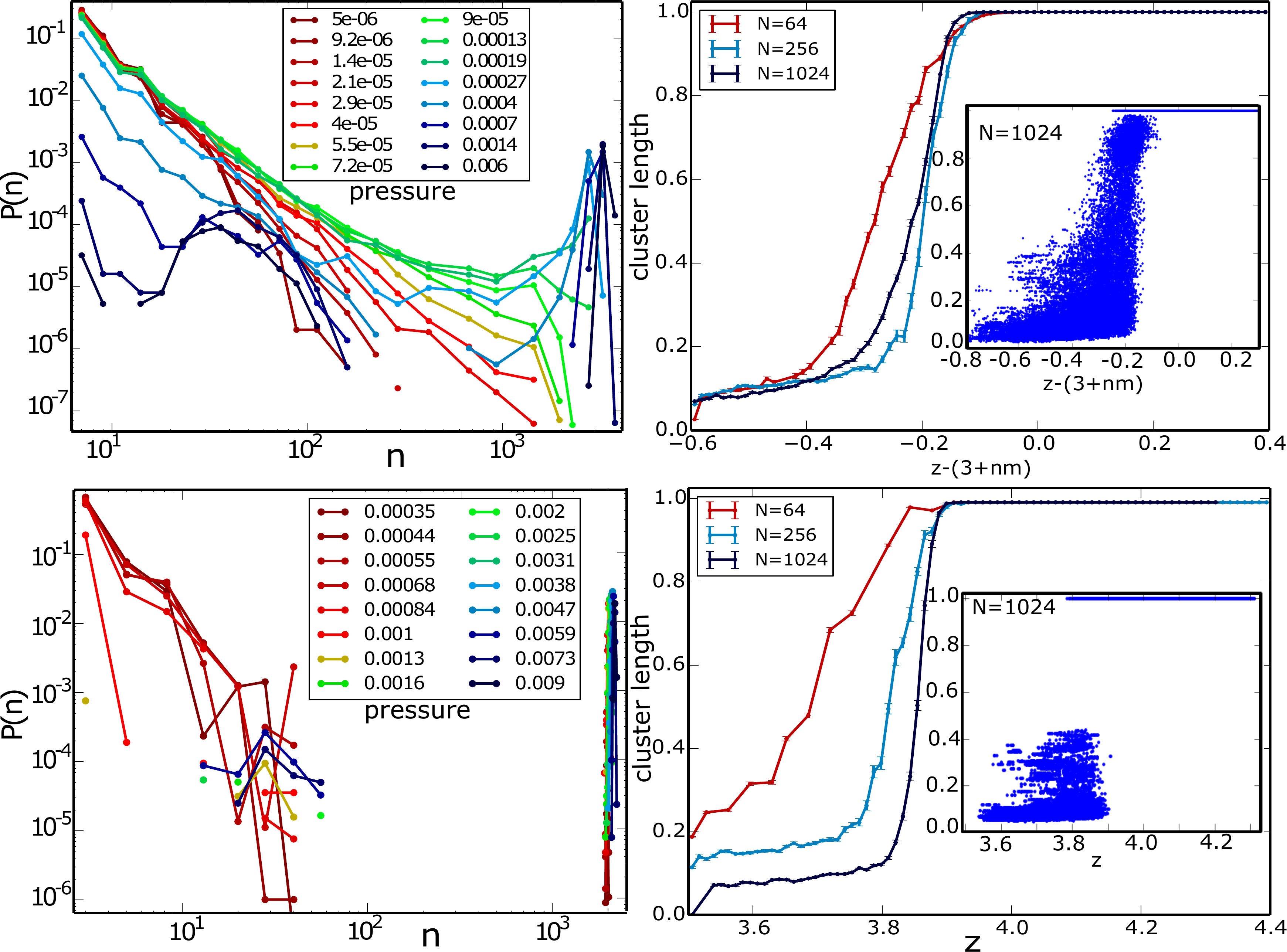}
 \caption{Rigid cluster size statistics. Left we show the cluster size distribution for pressures accross jamming, while on the right is the scaling of the largest cluster size with $dz$, for 3 system sizes. Top row: frictional systems, $N=64-1024$ , $\dot{\gamma}=10^{-6}$, 5 runs each at 8 packing fractions with $160000$ packings in the transition region. Bottom row: frictionless systems, idem with $64000$ samples at packing fractions $\phi=0.835-0.85$.}
  \label{fig:rigidity_transition}
   \vspace{-5mm}
\end{figure}

Fig. \ref{fig:Genziso_panel} demonstrates that $dz$ is an appropriate mean-field parameter. In light of Fig. \ref{fig:time_dependence}, we conclude that the spatial decomposition indeed plays a role in the macroscopic response of the system. We now ask how rigid cluster analysis can help uncover the nature of the frictional jamming transition.
In Fig. \ref{fig:rigidity_transition}, top row left, we show the rigid cluster size distribution $p(n)$, where $n$ is the number of bonds in a rigid cluster, across the frictional jamming transition. We observe a change characteristic of a \emph{second-order} phase transition, with a cluster size distribution which broadens approaching the transition, and then the emergence of a system-size percolating cluster above the transition. At the transition, we observe a power-law distribution with an exponent $\alpha\simeq-2.5$. For comparison, for connectivity percolation in two dimensions $\alpha=-187/81=-2.31$ \cite{Aharony85} and a self-organized rigidity percolation model yields $\alpha=-1.94$~\cite{Briere06}.
To help pinpoint the location of the transition, we plot the maximum cluster length $(L_x^2+L_y^2)^{1/2}/\sqrt{2}L$ against $dz$. It approaches unity near $dz=-0.15$ rather than $dz=0$, consistent with the picture of rigid and floppy regions coexisting in an overall rigid system emerging from Fig. \ref{fig:simulated_game}. So does an equivalent measure, the spanning probability (Fig. S7). Moreover, this downward shift survives in both the large $N$ (Fig.~\ref{fig:rigidity_transition} top right) and the $\dot{\gamma}\rightarrow0$ (Fig. S6) limit.

As comparison, we simulate a frictionless system across its frictionless jamming transition~\cite{Olsson07} with the same protocol as our frictional runs (Fig.~\ref{fig:rigidity_transition}, bottom row). Rotations are irrelevant to frictionless disks, so we use a standard $(2,3)$ pebble game here (see SI). In stark contrast to the second order transition discussed above, here we observe the hallmarks of a \emph{first-order} transition: The rigid cluster size distribution is either rapidly decaying at low pressures (most packings have no rigid clusters at all), or markedly bimodal without any intermediate-sized clusters for an order of magnitude. Also, the largest rigid cluster length is gapped (Fig. \ref{fig:rigidity_transition} right, inset), and depends strongly on system size.
This finding is consistent with recent results for frictionless systems where packings were found to be either fully rigid or fully floppy \cite{Ellenbroek15}. Again, we observe a downward shift in the transition point, though it is not as clear if it survives the large $N$ limit or the $\dot{\gamma}\rightarrow 0$ limit. Recent work demonstrates that the frictionless transition point can in fact change with protocol~\cite{Shattuck15}. 

\begin{figure}[t]
 \centering
 \includegraphics[width=0.99\columnwidth,trim=0mm 0mm 0mm 0mm,clip]{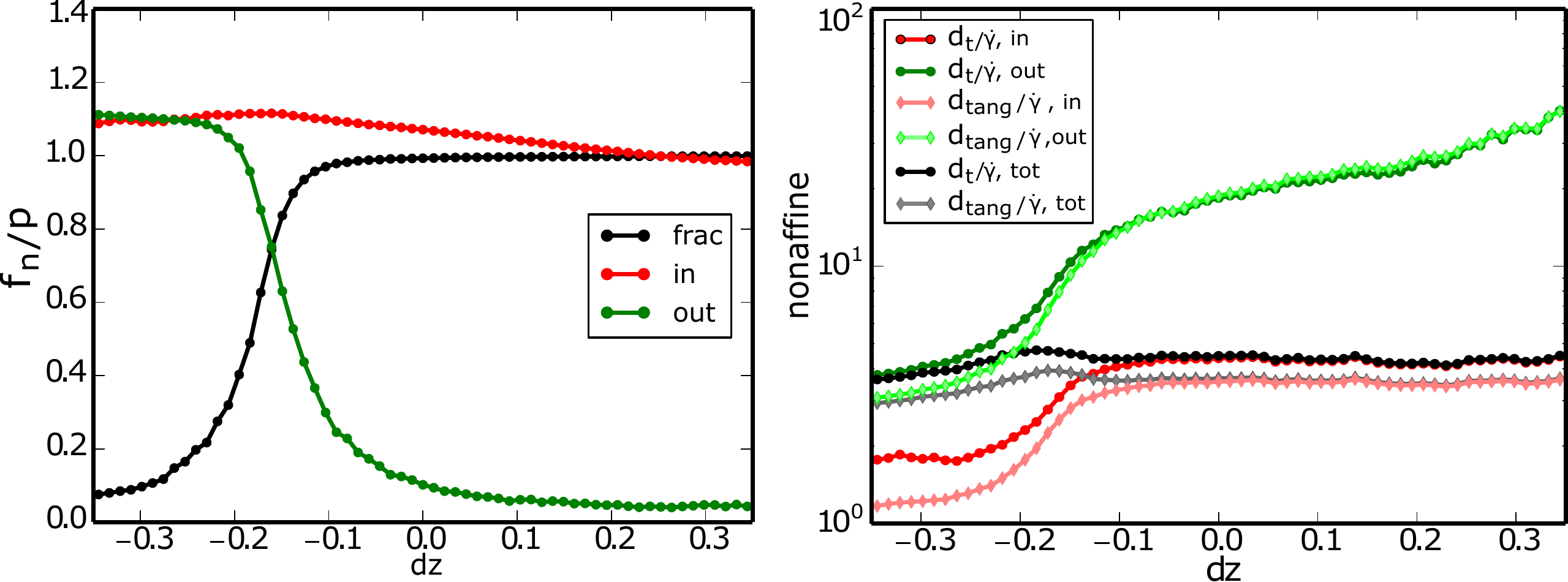}
 \caption{Correlation between rigid clusters and mechanical stability. \emph{Left:} Scaled normal force magnitude $f_n/p$ inside (red) and outside (green) the clusters, together with the rigid fraction of the system (black).  \emph{Right:} Nonaffine motion (tangential $d_t$ and contact slipping $d_{\text{tang}}$) inside and outside the cluster, and total.}
\label{fig:microscopics} 
\vspace{-0.5cm}
\end{figure}

To complete our analysis, we correlate rigid clusters with two other microscopic measures of rigidity for our frictional packings (see Fig. \ref{fig:microscopics}). First we measure the mean bond normal force scaled by the packing pressure $f_n/p$ over bonds belonging to either a rigid cluster or a floppy region, ensemble-averaged over packings with similar $dz$ (left). We see that $f_n$ inside the clusters always closely tracks the pressure, confirming that clusters indeed always bear stress. In contrast, floppy regions markedly depend on $dz$: Below the transition, they bear the same load as rigid regions, but then rapidly lose load once the percolating rigid structure forms. Finally, the unstable regions become isolated rattlers which bear no load. We interpret this again as viscous forces dominating the stresses below the transition, where no spanning cluster can bear the load. The gradual decrease of non-load bearing rattlers above jamming is well-known in static systems~\cite{Torquato}; we put it into a dynamical context here for the first time. Spatial correlations in the forces are also found in shear jamming with the transition occuring below the isostatic point~\cite{Bi11}.
The second measure are the nonaffine motions, which are known to dramatically increase approaching the frictionless~\cite{Ellenbroek06} and frictional~\cite{Henkes10} jamming transitions. We measure the relative tangential motion of the centers $\langle | d_t |\rangle$  and tangential sliding at the contacts $\langle | d_{\text{tang}}|\rangle$. Let $\mathbf{r}_{ij} = \mathbf{r}_j(t)-\mathbf{r}_j(t)$ be the vector linking two neighbouring particles' centers, $\hat{\mathbf{t}}_{ij}$ the tangential unit vector at the contact, $R_i$ the particle radii, and $\alpha_i$ their angles (see Fig. S1). Then \cite{Henkes10,Somfai07}
\begin{align}
 &d_t^{ij} =  \dot{\mathbf{r}}_{ij} \cdot \hat{\mathbf{t}}_{ij},
 &d_{\text{tang}}^{ij} = \dot{\mathbf{r}}_{ij}  \cdot \hat{\mathbf{t}}_{ij} - \left( R_i \dot{\alpha}_i +  R_j\dot{\alpha}_j \right).
\end{align} 
Normalized by the strain rate $\dot{\gamma}$, nonaffine motion is signalled by values above $0.5$. For both rigid and floppy regions, above and below the transition, motion is strongly nonaffine (Fig. \ref{fig:microscopics}, right, note log scale). However, displacements in floppy regions are much more nonaffine compared to rigid regions on both sides of the transition, culminating in values $d_t/\dot{\gamma}>20$ for isolated rattlers in rigid packings. The normal displacements remain at $d_n /\dot \gamma \approx 0.5$ throughout (see Fig. S3). Links between nonaffine buckling and local rigidity have been pointed out previously~\cite{Tordesillas}. We also observe a peak in the total nonaffine motion accross the transition.

{\it Conclusions:} 
In sum, we adapt ideas from rigidity percolation to characterize the frictional jamming transition of slowly sheared packings beyond mean-field level.
We show that while generalized isostaticity is a good mean-field criterion, spatial correlations do matter: a packing can be jammed below global isostaticity if it contains both floppy regions and a system-spanning rigid cluster, resembling spring networks in this regard.
The emergence of such a cluster appears to be first-order in frictionless packings, with a sudden jump from microscopic to system-wide clusters, but second-order in the frictional case, with a power law distribution of cluster sizes at the transition.
In particular, \emph{partial rigidity} is unique to \emph{frictional} packings.

The key instrument in obtaining those new results is rigid cluster decomposition. It allows us to draw connections between disordered spring networks, where it is the norm, and granular packings~\cite{Ellenbroek15,jamminggraph}.
By nature, cluster decomposition ignores the detail of contact forces, which may explain why it only partially correlates with local stresses: presumably, arching is only possible around \emph{small} floppy regions.
It does, however, account for the dynamical nature of fully-mobilized contacts, thus highlighting their central role in frictional packings. Applied strain leads to internal rearrangements, moving through a phase space of rigid, non-rigid and crucially partially rigid packings due to the second-order nature of the transition.
Incorporating this internal variance of accessible states through an internal field, Grob {\it et al.} were able to explain the phenomenology of hysteresis in frictional jamming~\cite{Grob14}. Our work begins to provide a microscopic basis for such phenomenology.

\begin{acknowledgments}
\emph{Acknowledgments.}
JMS acknowledges useful discussions with Wouter Ellenbroek and Martin van Hecke and NSF-DMR-1507938 for support.
\end{acknowledgments}

\bibliography{friction}

\end{document}


\graphicspath{{figures/}}

\title{Supplemental material}

\maketitle

\renewcommand{\thefigure}{S\arabic{figure}}
\setcounter{figure}{0}
 \renewcommand{\topfraction}{1.0}

\section{Generalized pebble game}

The foundation of extending ideas of constraint-counting beyond mean-field in two dimensions is Laman's theorem~\cite{Laman70}. Specifically, it states that a network with $N$ vertices
is generically, minimally rigid in two dimensions if and only if
it has $2N-3$ bonds and no subgraph of $n$ vertices has more than
$2n-3$ bonds. Here minimal rigidity signifies that should any one contact/bond be deleted from the packing then the system breaks up into at least two rigid clusters, i.e. there are no redundant contacts and no floppy modes (prior to the deletion). Rigid clusters are defined along contacts/bonds and consist of connected bonds that are rigid with respect to each other in the sense that there is a nonzero energy cost to moving one bond with respect to another.  In the language of jamming, the existence of a spanning cluster that is minimally rigid translates to the packing being precisely at the jamming transition point. Laman's theorem is purely combinatorial. The power of the theorem is that it goes beyond simple constraint counting, which is just a necessary condition, and additionally accounts for whether or not the constraint is redundant since such contacts do not contribute to the emergence of rigidity.  
 
The $(k=2,l=3)$ pebble game ~\cite{Jacobs97} is an efficient algorithm to implement Laman’s theorem. It also provides us with a method for identifying rigid clusters in the packing. In the $(2,3)$ pebble game, $k=2$ pebbles, corresponding to the local degrees of freedom for each particle, are placed on each node/particle of the packing/network~\cite{Jacobs97}. These pebbles can be moved along the contacts in the contact network in such a way to explore all possible subgraphs. If a contact is indeed an independent constraint, then a pebble will be assigned to that contact. The pebble game stops when there are $l=3$ free pebbles (not assigned to any contact), corresponding to the $l=3$ global degrees of freedom, which do not tell us about the nontrivial deformations of the packing. 

While the original work on the pebble game focused on the $k=2,l=3$ case~\cite{Jacobs97}, Lee and Streinu addressed the general case~\cite{Lee2008}. Here is a description of the general case based on their work~\cite{Lee2008}. To determine if the contact network is under or over or minimally constrained, the rules are as follows: (1) No more than $k$ pebbles are allowed on any nodes in the contact network, and (2) a bond is labeled as an independent constraint when there are at least $l+1$ pebbles in total present at the two nodes connected by the contact.  Allowable pebble moves are: (1) Pebbles may be moved from node to node by searching a directed graph such that if a pebble is found then the bonds along the directed path are reversed and (2) once a pebble has been assigned to a bond, the direction of that bond is determined by the node the pebble came from. The game is then played until all the bonds have been processed. If there are $l$ free pebbles left over and no unassigned bonds, then the contact network is minimally rigid, or at the jamming transition. If there are more than $l$ pebbles left over, the network is floppy. If there are unassigned bonds with $l$ pebbles left over, then network is overconstrained. 

\begin{figure}[t]
 \centering
  \includegraphics[width=0.99\columnwidth,trim=0mm 0mm 0mm 0mm,clip]{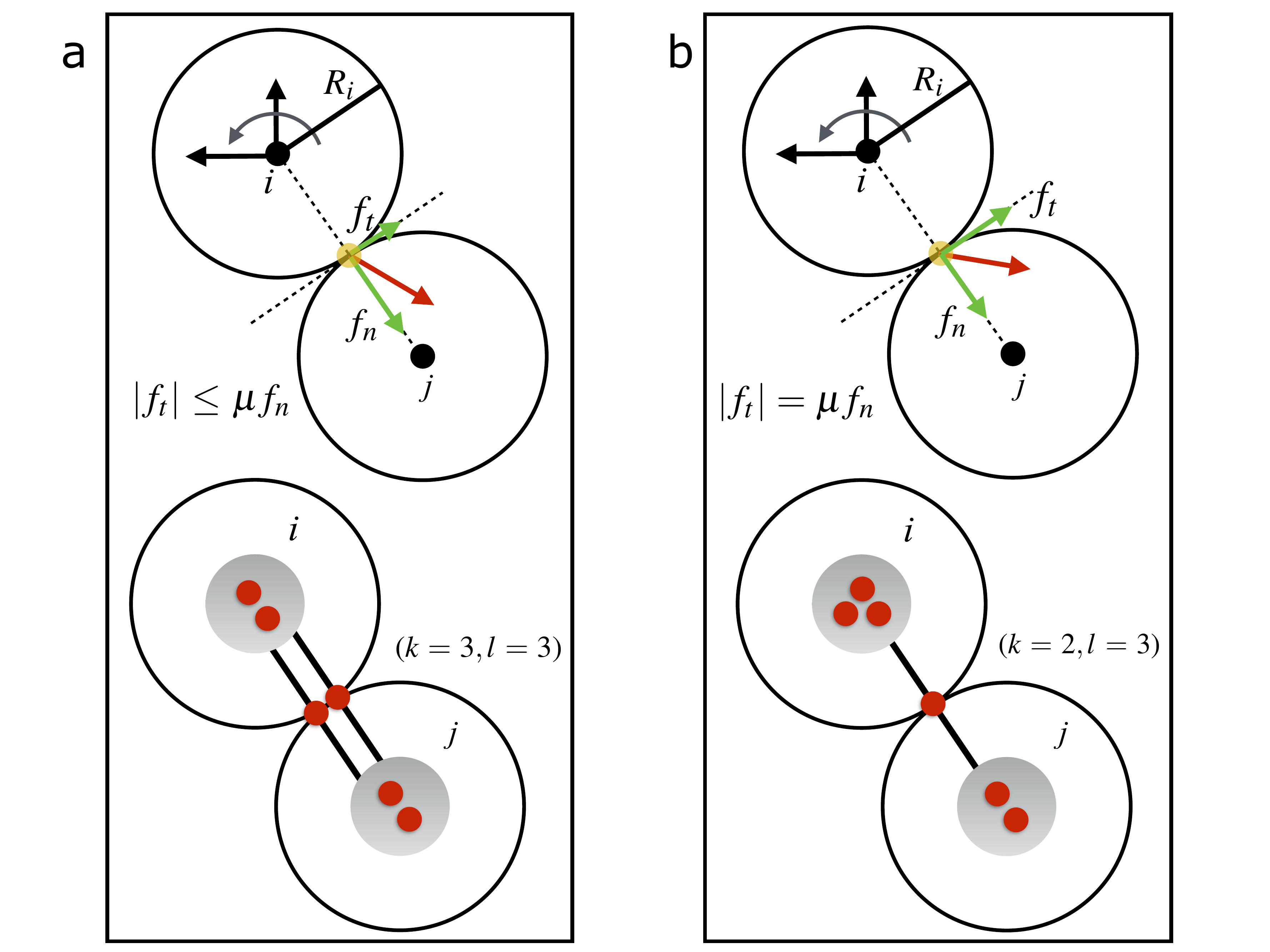}
 \caption{(a) Top: Schematic illustrating the contact geometry for frictional contacts, the force on particle $i$ due to particle $i$ assuming particle $j$ is stationary, and the displacement and rotation of particle $i$. Bottom: Mapping to constraint network. (b) Same as (a) except for sliding contacts.  }
  \label{fig:force_schematic}
\end{figure}

The pebble game can be extended to compute rigid clusters by temporarily pinning $l$ free pebbles and then marking two particles in contact as rigid. Neighboring disks in contact are then explored to free a pebble. If a free pebble is found, the neighboring disks and their associated contacts are flexible. If the converse, the associated contacts are rigid. Rigid clusters are defined along contacts/bonds and consist of connected contacts that are rigid with respect to each other. This procedure can then be used to map out the rigid clusters. A system that is not jammed will not have a spanning rigid cluster, so the existence of a spanning rigid cluster determines the state of the system (fluid or solid). The existence of a spanning minimally rigid cluster determines the transition point.  

\begin{figure}[t]
 \centering
  \includegraphics[width=0.75\columnwidth,trim=50mm 30mm 50mm 25mm,clip]{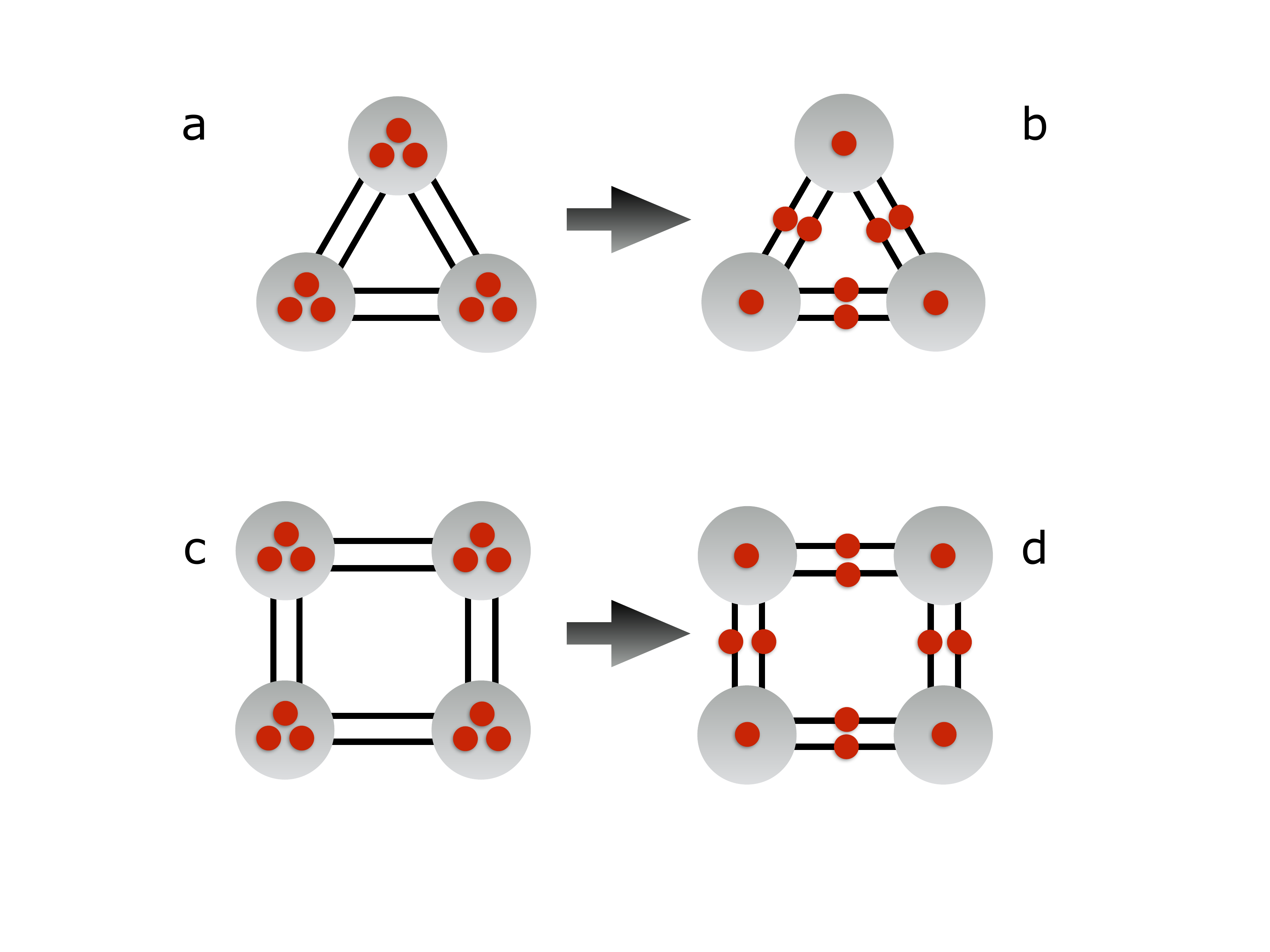}
 \caption{The (3,3) pebble game for two sample configurations. Top row: Three particles connected by frictional bonds become a rigid cluster, since there are a total of 3 pebbles left. Bottom row: Four particles in a square configuration are flexible, since there are a total of 4 pebbles left.}
  \label{fig:pebble_game}
\end{figure}

In order to identify rigid clusters in frictional packings, we first must extend the commonly used ($k=2,l=3$) pebble game for frictionless disks, where only normal (central) forces exist, to frictional disks, where both normal forces and tangential forces exist.  Since there are now two translational degrees of freedom and one rotational degree of freedom for each particle, $k=3$. The constraints can be taken into account as explained above, i.e. once a pebble is assigned to a contact in the pebble game, then that contact is considered an independent constraint.  However, to account for the additional constraints due to tangential forces in the frictional case, we introduce a second bond for each frictional contact into what we now call the ``constraint network''. The pebble game then explores the constraint network to see if that additional rotational degree of freedom can be independently constrained. This second bond in the constraint network is only added to frictional contacts below the Coulomb criterion, where the normal and tangential forces are independent of each other.  For fully mobilized contacts (contacts at the Coulomb criterion), the tangential and normal forces are no longer independent such that only one bond in the constraint network is needed. Since the number of global degrees of freedom in two dimensions is still 3, we arrive at a ($k=3,l=3$) pebble game where contacts below the Coulomb criterion are denoted as double bonds in the constraint network and contacts at the Coulomb criterion are denoted as single bonds in the constraint network. While the $(3,3)$ pebble game has been used to analyze the rigidity in body-bar networks~\cite{Tay1984}, it has not yet been used for frictional disks. See Fig.~\ref{fig:force_schematic}.

As a test of this approach to identifying rigid clusters in frictional disk packings, one can consider a few small packings with three or four disks (see Fig. \ref{fig:pebble_game}). In the frictionless case, the smallest nontrivial minimally rigid packing (and corresponding constraint network) is a triangle. In the frictional case, if we consider a triangle with each contact below the Coloumb criterion, then there are two bonds between each node in the constraint network (panel a). After performing the ($2,3$) pebble game, there are 3 free pebbles left and so the constraint network is minimally rigid (panel b).  Note that this result is in keeping with three gears in a triangle not being able to turn. And if at least one of the ``double bonds'' in the constraint network is removed from the triangle, then the packing is flexible since sliding between two of the disks is possible. If instead we consider four frictional disks packed as a square with each contact below the Coloumb criterion (panel c), then there are four free pebbles left after the ($3,3$) pebble game is played and the packing is flexible just as four gears arranged in a square can turn (panel d).

\begin{figure}[h]
 \centering
  \includegraphics[width=0.99\columnwidth,trim=0mm 0mm 0mm 0mm,clip]{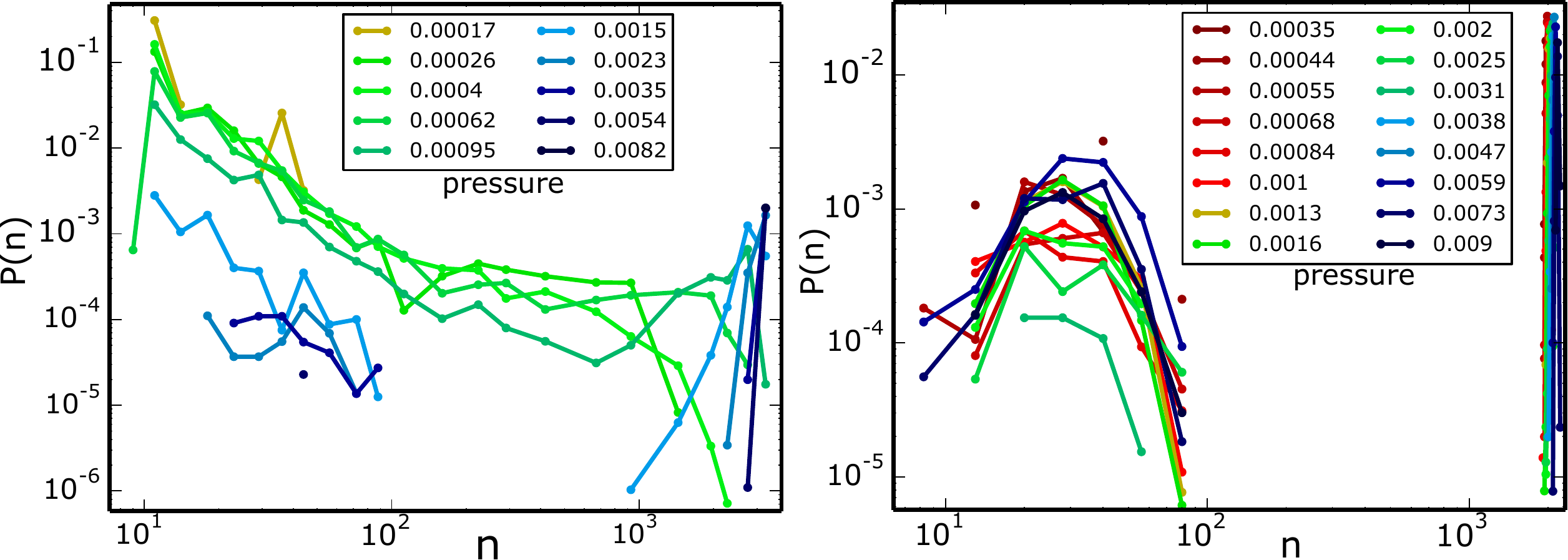}
 \caption{Cluster size distributions for the (3,2) and (2,2) pebble games for frictional and frictionless systems, respectively. Left: frictional system at $\dot\gamma=10^{-6}$ and $\zeta=1.0$. Right: frictionless system, same parameters as Figure 4. Despite the limited statistics, the frictional transition still appears continuous, while the frictionless transition retains a gap in cluster sizes.}
  \label{fig:pebble_3_2_and_2_2}
\end{figure}

Finally, we address boundary conditions.
For two-dimensional systems with periodic boundary conditions, global rotations are forbidden by symmetry and there are only \emph{two} global degrees of freedom (the two translations).
Clusters smaller than system size do not suffer this restriction, but a spanning cluster can only have two, rather than three, left-over pebbles.
In practice and for our purpose, there is little difference between the $l=2$ and the $l=3$ pebble game. Although some results are affected quantitatively, the location of the transition point remains the same and our main conclusion is unaffected: Frictional systems have a continuous cluster size distribution (Figure \ref{fig:pebble_3_2_and_2_2}, left), while frictionless systems show a distinct gap (Figure \ref{fig:pebble_3_2_and_2_2}, right).
In other words, the precise form of the pebble game does not affect the observed order of the transition.

\section{Additional results}
In this section, we present a series of supplementary results which complement our message of frictional jamming as a rigidity transition, by varying the strain rate $\dot{\gamma}$, the friction coefficient $\mu$ and the viscous damping coefficient $\zeta$, and by presenting alternative measures of the transition.

\begin{figure}[b]
 \centering
  \includegraphics[width=0.99\columnwidth,trim=0mm 0mm 0mm 0mm,clip]{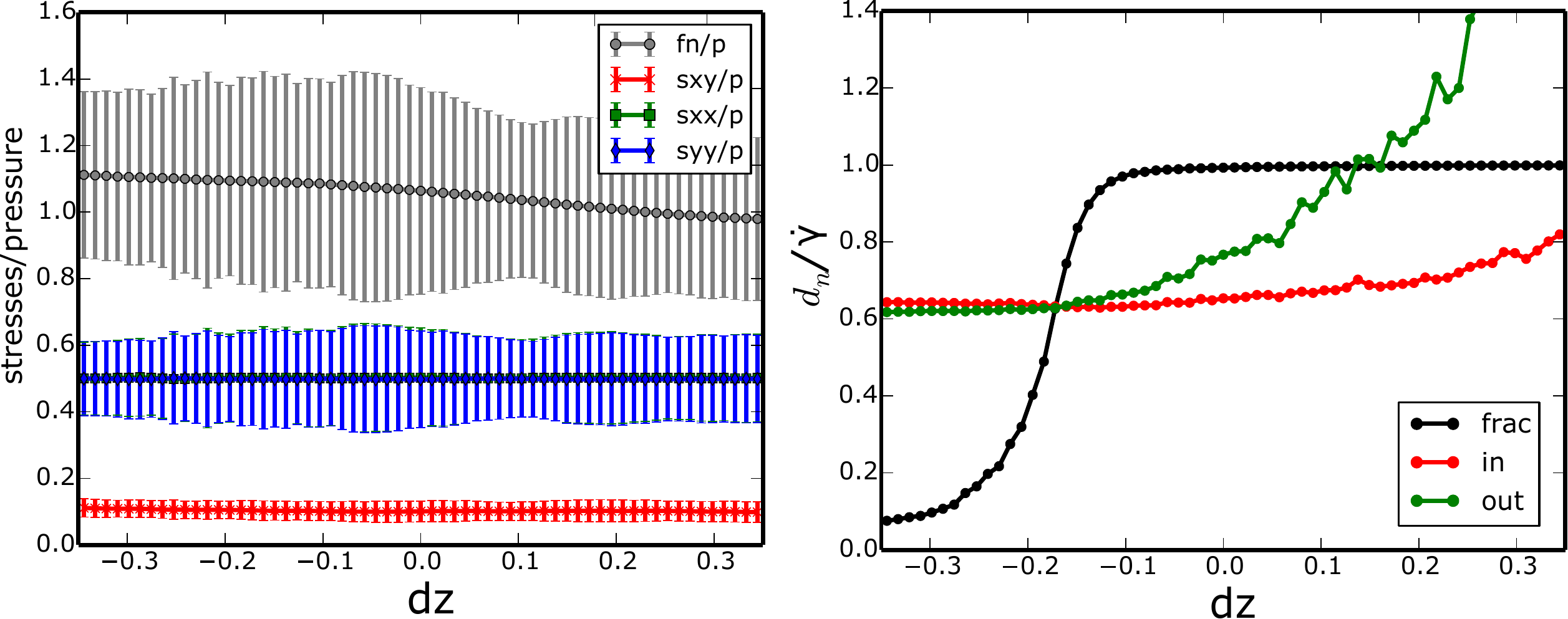}
 \caption{Right: Ratio of other stress components to pressure: normal forces, xy, xx and yy components of the stress tensor. There are no systematic variations except for $f_n$, where the mean contact number has an influence. The bars indicate the standard deviation of the fluctuations, they are not error bars of the mean. Right: Normal part of the decomposition of the motion inside and outside the cluster. The normal displacements are close to the affine prediction $d_n /\dot \gamma =0.5$. }
  \label{fig:stress_nonaffine}
\end{figure}

\emph{Stress measures.} We begin, however, by confirming that our use of the pressure as a marker of jamming is consistent with other possible definitions, in particular the shear stress $\sigma_{xy}$. Figure \ref{fig:stress_nonaffine} shows the ratio of four alternative measures of the stresses and the pressure (normal force $f_n$, shear stress $\sigma_{xy}$, diagonal stress components $\sigma_{xx}$ and $\sigma_{yy}$). We have indicated the standard deviation of the fluctuations by error bars. Except for $f_n$, where the mean contact number matters, this ratio is virtually independent of $dz$, with a finite and constant fluctuation amplitude, indicating that all of these quantities are equivalent measures of the distance to the transition.

\emph{Nonaffine motion - normal displacements.} Complementing Figure 5 in the main text, Figure \ref{fig:stress_nonaffine} (right) shows the normal components of the local displacments, $ d_n^{ij} =  \dot{\mathbf{dr}}_{ij} \cdot \hat{\mathbf{n}}_{ij}$, scaled by the strain rate. The values both inside and outside the clusters are close to the affine prediction, $d_n/\dot\gamma =0.5$.

\begin{figure}[h!]
 \centering
  \includegraphics[width=0.99\columnwidth,trim=0mm 0mm 0mm 0mm,clip]{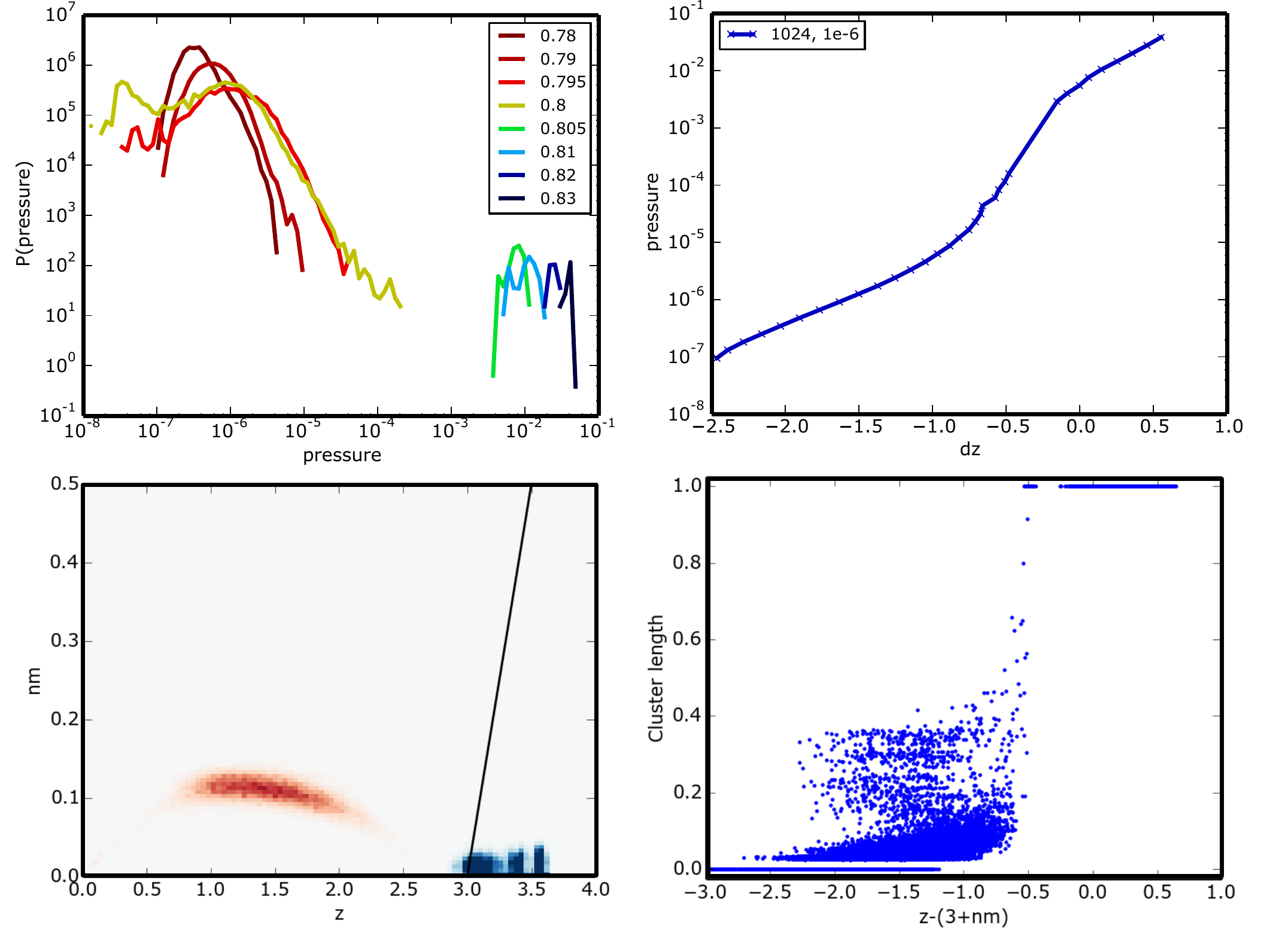}
 \caption{Properties of a frictional system with friction coefficient $\mu=1$. Top left: pressure distributions for $8$ values of $\phi$ accross the jamming transition. Top right: pressure - dz scaling, showing the same collapse as for $\mu=0.1$. Bottom left: Two dimensional histogram of states, colored according to jammed (blue) and unjammed (red); criterion for jamming $p\geq 10^{-3}$. Bottom right: Scatter plot of the maximum cluster size as a function of $dz$.}
  \label{fig:friction_mu1}
\end{figure}

\emph{Influence of $\mu$}. In Figure \ref{fig:friction_mu1}, we present equivalent data to the main article for $\mu=1$, i.e. in the high friction limit as opposed to the $\mu=0.1$ low friction limit. From the generalized isostaticity point of view, the main influence of the higher $\mu$ is a smaller number of fully mobilized contacts, consistent with~\cite{Shundyak07,Henkes10}. In the $n_m$--$z$ diagram (panel c), jammed configurations have close to no sliding contacts, and rest just within the stable region. Unjammed configurations have more sliding contacts, and also significantly fewer contacts overall.
As shown in panel a, the jamming transition shifts to lower $\phi$ - from our limited resolution it lies in the range $0.8 < \phi_J < 0.805$. Unlike for $\mu=0.1$ in Figure 3, we do not observe runs which flip between jammed and unjammed states, possibly since we are not close enough jamming. If we plot pressure vs. $dz$, our data collapse on nearly an identical curve to the equivalent $N=1024$, $\gamma=10^{-6}$ line in Figure 2.
Finally, we have repeated the rigid cluster analysis for $\mu=1$ (panel d). Though our statistics is limited,  we can clearly see a growing cluster when approaching the transition from below, and then a spanning cluster appears somewhat below $dz=0$.

\begin{figure}[h!]
 \centering
  \includegraphics[width=0.99\columnwidth,trim=0mm 0mm 0mm 0mm,clip]{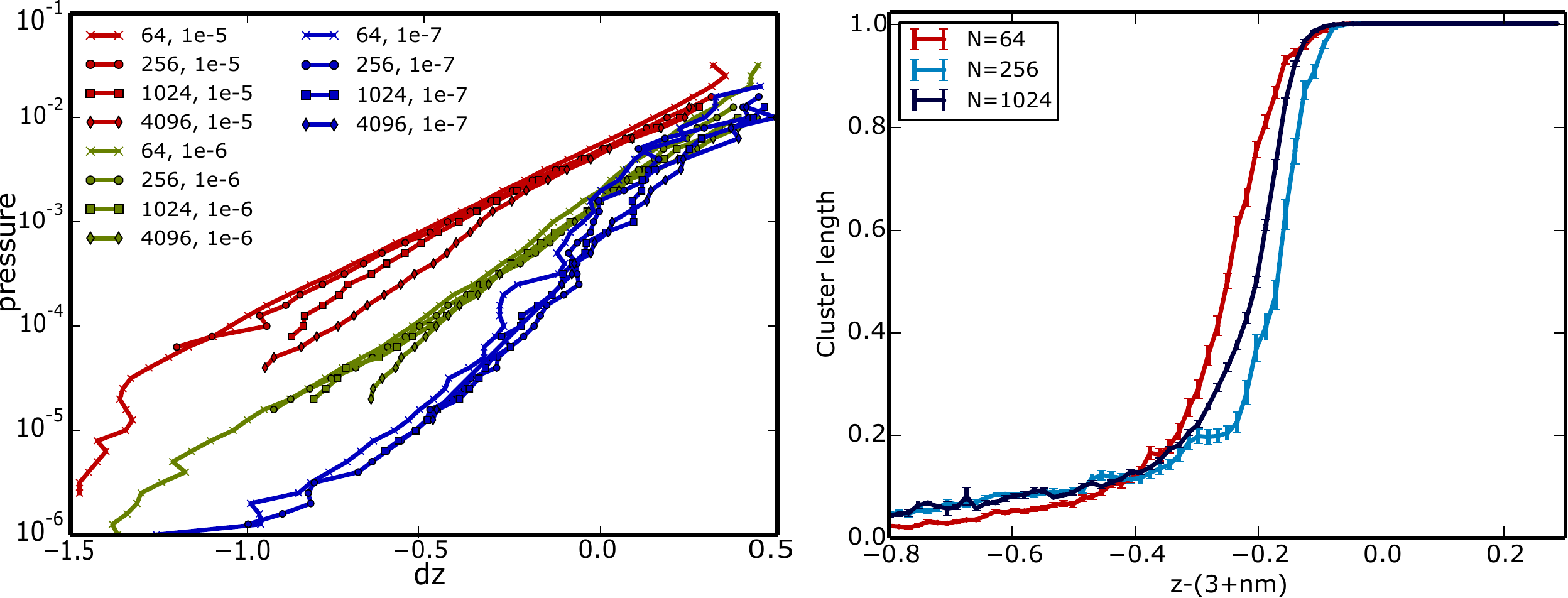}
 \caption{Results for higher viscous friction values $\zeta=1.0$. Left: pressure - dz plots; the differences compared to the low friction case $\zeta=0.1$ are at unjammed dz, where there are larger contributions to the viscous stresses. Right: Rigidity transition scaling for three different system sizes. The result is essentially unchanged from Figure 4.}
  \label{fig:xi_1}
\end{figure}

\emph{Influence of $\zeta$.} Figure \ref{fig:xi_1}. The viscous damping coefficient $\zeta$ chiefly affects the viscous part of the forces, which is included in our stress measurements. 
Intuitively, the mechanism of the transition should not depend on $\zeta$ for low enough strain rate since the viscous forces $f_v^{\zeta} \sim \zeta (\mathbf{v}_j-\mathbf{v}_i) \sim \zeta \dot{\gamma} \:n_{\text{aff}}$ go to zero (the last term is the nonaffine parameter). 
We find that while the value of $\zeta$ strongly increases the pressure values below jamming, consistent with the viscous forces dominating this regime, the pressure in the jammed region is virtually unaffected (left). 
The transition curve (right) strongly resembles the $\zeta=0.1$ case in Figure 4, albeit somewhat smoothened out.

\begin{figure}[h!]
 \centering
  \includegraphics[width=0.99\columnwidth,trim=0mm 0mm 0mm 0mm,clip]{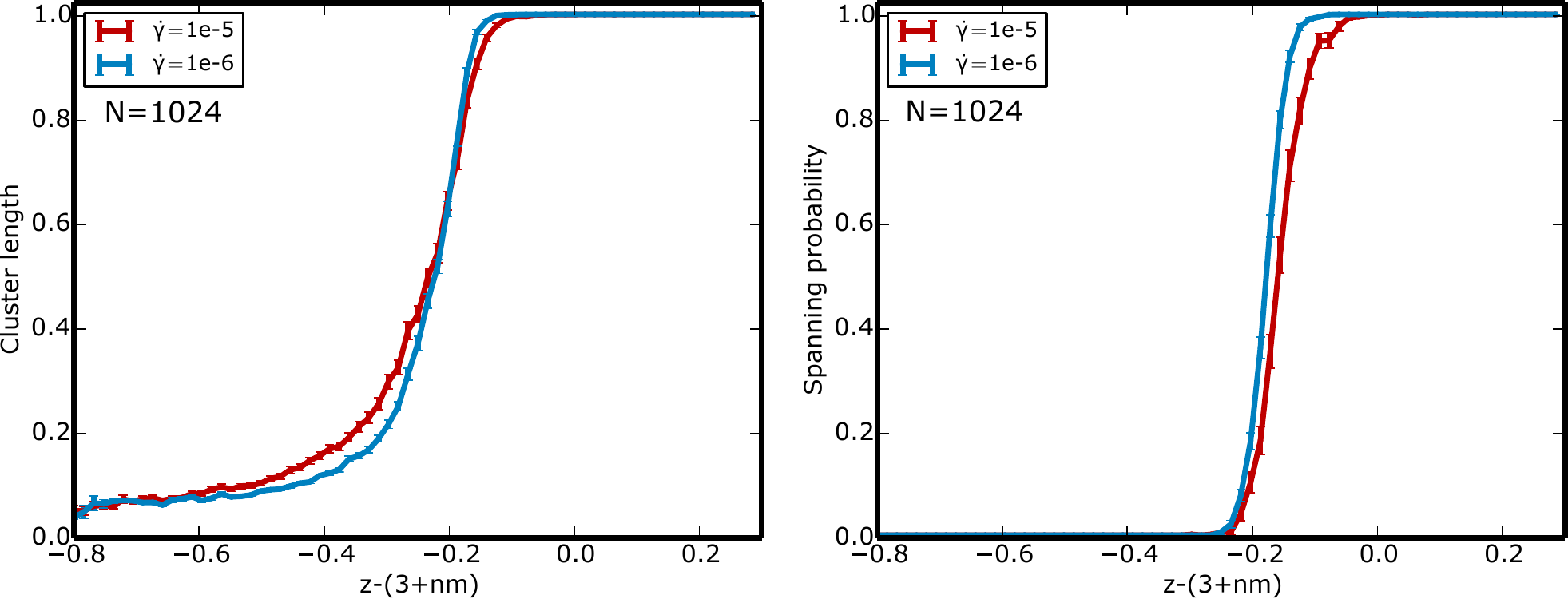}
 \caption{Rigidity transition properties of the frictional system as a function of strain rate: for decreasing strain rates, the transition becomes sharper. Left: length of the largest cluster. Right: Spanning probability. All systems were simulated for $N=1024$ and $\zeta=0.1$.}
  \label{fig:strainvary}
\end{figure}

\emph{Strain rate dependence.} While part of the $\dot{\gamma}$ dependence was explored in Figure 2, we compare the transition curves for $\dot \gamma = 10^{-5}$ and $\dot \gamma = 10^{-6}$ in Figure \ref{fig:strainvary}. We observe a somewhat narrower transition for the lower strain rate, consistent with a sharp transition as $\dot \gamma \rightarrow 0$. (Not shown: Measures of the nonaffine parameter are consistent with this slight broadening at $\dot \gamma = 10^{-5}$). Unfortunately, reliable $N=1024$, $\dot \gamma = 10^{-7}$ results are beyond our current computational capacities, but results for smaller $N$ point to a further narrowing.

\begin{figure}[h!]
 \centering
  \includegraphics[width=0.99\columnwidth,trim=0mm 0mm 0mm 0mm,clip]{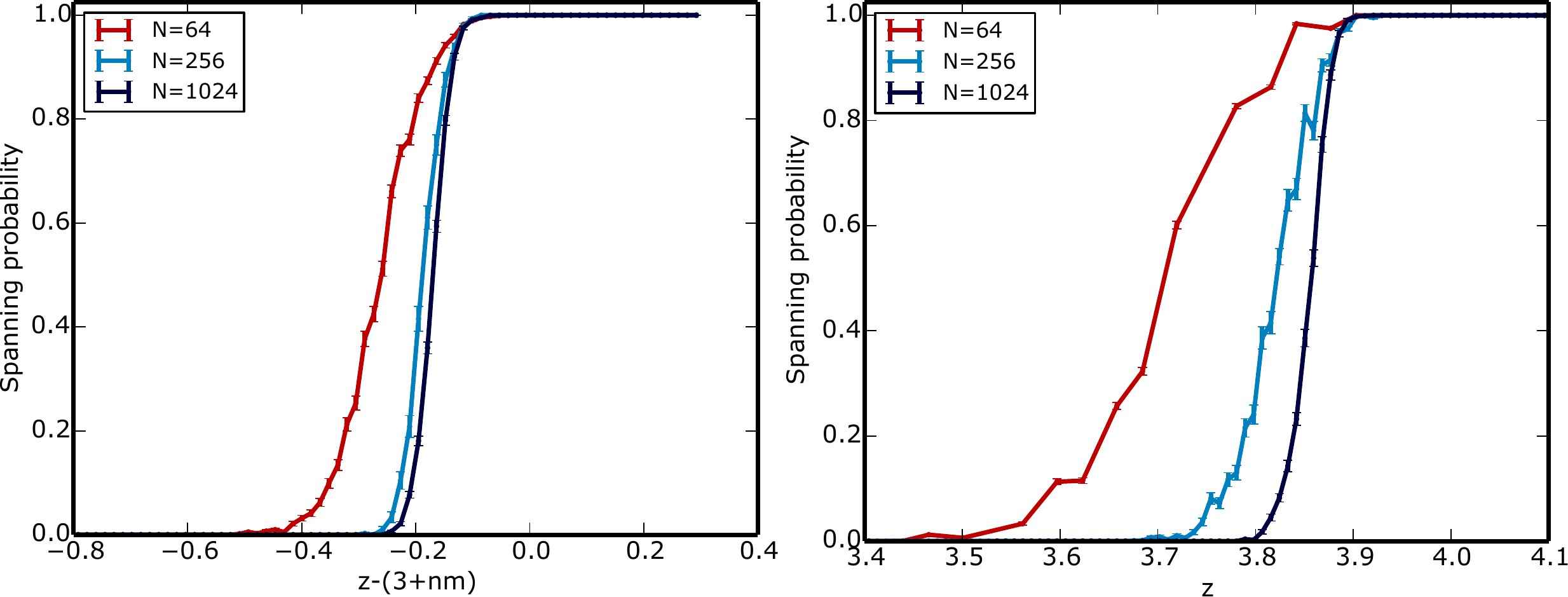}
 \caption{Probability of a spanning cluster for frictional (left) and frictionless (right) systems, same data set as Figure 4.}
  \label{fig:spanning_probability}
\end{figure}

\emph{Other measures of rigidity.} In rigidity percolation theory, an important measure is the \emph{spanning probability} of the system, i.e. the probability of the existence of a system-size cluster. This allows us to then pinpoint the unique transition point in the limit $N\rightarrow \infty$. In Figure \ref{fig:spanning_probability}, we show the spanning probablities for three system sizes for both frictional (left) and frictionless (right) systems. For both, we observe a dramatic narrowing of the transition range with $N$. For both, we also confirm that the transition point is in fact \emph{below} the isostaticity point, at about $dz=-0.15$ for frictional systems and $z=3.9$ for frictionless ones. In Figure \ref{fig:strainvary} (right), we also show the dependence of the spanning probability on $\dot\gamma$, again showing that higher strain rates slightly smooth out the transition.

\bibliography{friction}